\documentclass[superscriptaddress,reprint,amsmath,amssymb,aps]{revtex4-2}
\usepackage[english]{babel}
\usepackage{amsmath,amssymb,mathtools,float}
\usepackage{graphicx}
\usepackage{dcolumn}
\usepackage{bm}
\usepackage{xr}
\usepackage[version=4]{mhchem}
\usepackage{hyperref}
\usepackage[dvipsnames]{xcolor}
\usepackage{siunitx}
\DeclareSIUnit\angstrom{\text {Å}}
\DeclareSIUnit\molar{\text {M}}
\usepackage{layouts}
\usepackage{mathtools} 
\usepackage{amssymb}   
\usepackage{siunitx}   
\usepackage{soul}
\usepackage{float}

\usepackage{tikz}

\usepackage{tabularx}   
\usepackage{longtable} 
\usepackage{xltabular} 
\usepackage{booktabs}   

\usepackage[capitalize]{cleveref}

\makeatletter

\begin{document}

\title{Electrostatics slows down the breakup of liquid bridges on solid surfaces}

\author{Salar Jabbary Farrokhi}
\thanks{S.J.F. and A.D.R. contributed equally to this work.}
\affiliation{
 Institute for Nano- and Microfluidics, TU Darmstadt, 
 Peter-Gr\"unberg-Stra{\ss}e 10, D-64287 Darmstadt, Germany
}

\author{Aaron D. Ratschow}
\thanks{S.J.F. and A.D.R. contributed equally to this work.}
\affiliation{
 Institute for Nano- and Microfluidics, TU Darmstadt, 
 Peter-Gr\"unberg-Stra{\ss}e 10, D-64287 Darmstadt, Germany
}
 
\author{Steffen Hardt}
\email{hardt@nmf.tu-darmstadt.de}
\affiliation{
 Institute for Nano- and Microfluidics, TU Darmstadt, 
 Peter-Gr\"unberg-Stra{\ss}e 10, D-64287 Darmstadt, Germany
}

\date{October 21, 2024}

\begin{abstract}

We experimentally study the breakup of water-glycerol liquid bridges on non-conductive surfaces and find that spontaneous charge deposition at the receding contact line, slide electrification, can have a substantial influence. Electrostatic forces slow down the dynamics during, and cause spontaneous motion of satellite drops after the bridge breakup. We show that our experimental observations align with slide electrification theory. 
Our findings demonstrate that slide electrification plays an important role in dewetting beyond drop-related scenarios. 
\end{abstract}

\maketitle

\section{Introduction}
\label{sec:headings}

The breakup of capillary liquid bridges between two parallel surfaces, referred to as \emph{free liquid bridges}, has been studied for several decades \cite{Eggers.1993,Papageorgiou.1995,Papageorgiou.1995b,Stone.1996,Brenner.1996,Eggers.2008,Eggers.2012}
and has important implications for printing processes like offset and inkjet printing \cite{Darhuber.2001,Lohse.2022}. In free liquid bridges, the dynamic breakup behavior is governed by the balance of three forces: the capillary force, which drives the dynamics, as well as viscous and inertial forces, which oppose the capillary force. It can be subdivided into three regimes: viscous, viscous-inertial, and inertial, depending on which is the main contribution opposing the driving force. In the viscous-inertial regime, viscous and inertial forces are of similar magnitude \cite{Li.2016}. Scaling relations and similarity solutions have been derived for all three regimes \cite{Papageorgiou.1995,Eggers.2012,Brenner.1996}. 
In contrast, the breakup of capillary liquid bridges on solid surfaces, \emph{wetting liquid bridges}, has largely remained obscure. For example, wetting liquid bridges form on surfaces with nonuniform wettability and are especially important for applications like multilayer inkjet printing, as is used in the production of electronic circuits, fuel cells, and solar panels \cite{Lohse.2022}. Upon evaporation, wetting liquid bridges become unstable and break up \cite{Hartmann.2019}.
In contrast to free liquid bridges, wetting liquid bridges are not axisymmetric but are bounded by a three-phase contact line. This introduces contact line forces as an additional contribution governing the breakup dynamics.  
For low-viscosity wetting liquid bridges, the breakup is governed by a balance of capillary and inertial forces \cite{Hartmann.2021}. However, at higher viscosities, as commonly found in the above mentioned applications, contact line friction often dominates \cite{Farrokhi.2024}. Consequently, the solid substrate properties have a substantial impact on the breakup of viscous wetting liquid bridges.

In dynamic wetting, besides viscous dissipation \cite{Voinov.1977,Cox.1986} and molecular processes at the contact line \cite{Blake.1969,Blake.2002}, surface roughness, heterogeneties, defects \cite{Raphael.1989}, elastic deformation \cite{Lester.1961}, and adaptation \cite{Butt.2018} can lead to additional forces hindering the motion of the contact line.
Recent work on sliding and impacting drops shows that in addition to these, electrostatic phenomena contribute to the forces acting on sliding and impacting drops  \cite{Li.2022,Li.2023,Wong.2022,Jin.2022,Bista.2024,Ratschow.2024}. This is due to an effect called slide electrification, i.e. the spontaneous deposition of charges by moving contact lines.

Here, we experimentally study the breakup of viscous wetting liquid bridges on conductive and non-conductive substrates and find that the breakup is significantly slower on non-conductive substrates. We systematically trace the effect back to slide electrification and correlate parametric dependencies with corresponding models. Our work suggests that the effects of slide electrification do not only concern drops on surfaces, but are generally important in wetting processes with receding contact lines. 

 \section{Experiments}

\begin{figure}
    \centering
    \includegraphics[width=6.5cm,page=1]{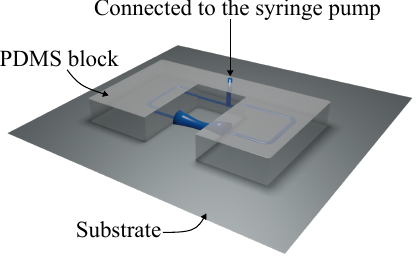}\label{Fig1}
    \caption{Schematic representation of the experimental setup with the microfluidic components. } \label{Fig1}
\end{figure}

The experimental setup, procedure, liquids, and substrates used in this study are similar to those in our previous study. For more details, the reader is referred to \cite{Farrokhi.2024}. However, a brief summary of the methodology is provided here. To form and break up a liquid bridge on a solid surface, a microfluidic setup, schematically depicted in \cref{Fig1}, is used. 
The corresponding PDMS structure is fabricated using a soft lithography protocol. It consists of two channels with semi-circular cross-sections, each having a diameter of \SI{0.5}{mm}. The PDMS block is placed on a substrate, and liquid is pumped through the channels by a syringe pump (KDS Legato 200) to form the liquid bridge on the substrate in the gap between the channels. 
Subsequently, the same syringe pump is used to withdraw the liquid and reduce the volume of the liquid bridge, leading it to an unstable configuration where it breaks up due to the well-known Rayleigh-Plateau instability. The evolution of the liquid bridge is recorded using a high-speed camera (Photron FASTCAM SA-1.1). An in-house MATLAB script is used to determine the minimum width of the capillary bridge and the corresponding velocity from the recorded images. 
Silicon wafers (CZ-Si wafer, 4-inch diameter, thickness 500 $\pm$ 50 µm,
Microchemicals GmbH, Germany) coated with thin layers of chromium (\SI{15}{nm}) and gold (\SI{50}{nm}) using the E-beam method (Balzers BAK 600) are chosen as the conductive substrates. Glass slides (soda lime, $76\times\SI{52}{mm}$, thickness $\SI{1}{mm}$, Paul Marienfeld GmbH, Germany) are used as the non-conductive substrates. 
A PDMS pseudo-brush coating is applied to the surfaces, resulting in advancing and receding contact angles of  approximately 109 and 95 degrees for water (Krüss DSA 100), respectively. 
The contact angles are the same for the two substrates.
Atomic force microscopy scans reveal that the root mean square roughness for both substrates is between 0.4 and \SI{0.6}{nm}.
Water (purified using a Milli-Q device; specific resistance \SI{18.2}{\mega \ohm \cdot cm} at \SI{25}{\degreeCelsius})
and glycerol ($>99.5$\%, Sigma-Aldrich, Germany)  mixtures are used as the working liquids, with glycerol contents ranging from 0\% to 70\%, corresponding to viscosities of 1 to \SI{23.7}{mPa\cdot s}. Throughout this article, we will refer to the liquids by their glycerol percentage (mass of glycerol / mass of the solution $\times$100).

\section{Results and Discussion}
\subsection{Satellite droplet motion}

\begin{figure*}
    \centering
    \includegraphics[width=0.75\textwidth]{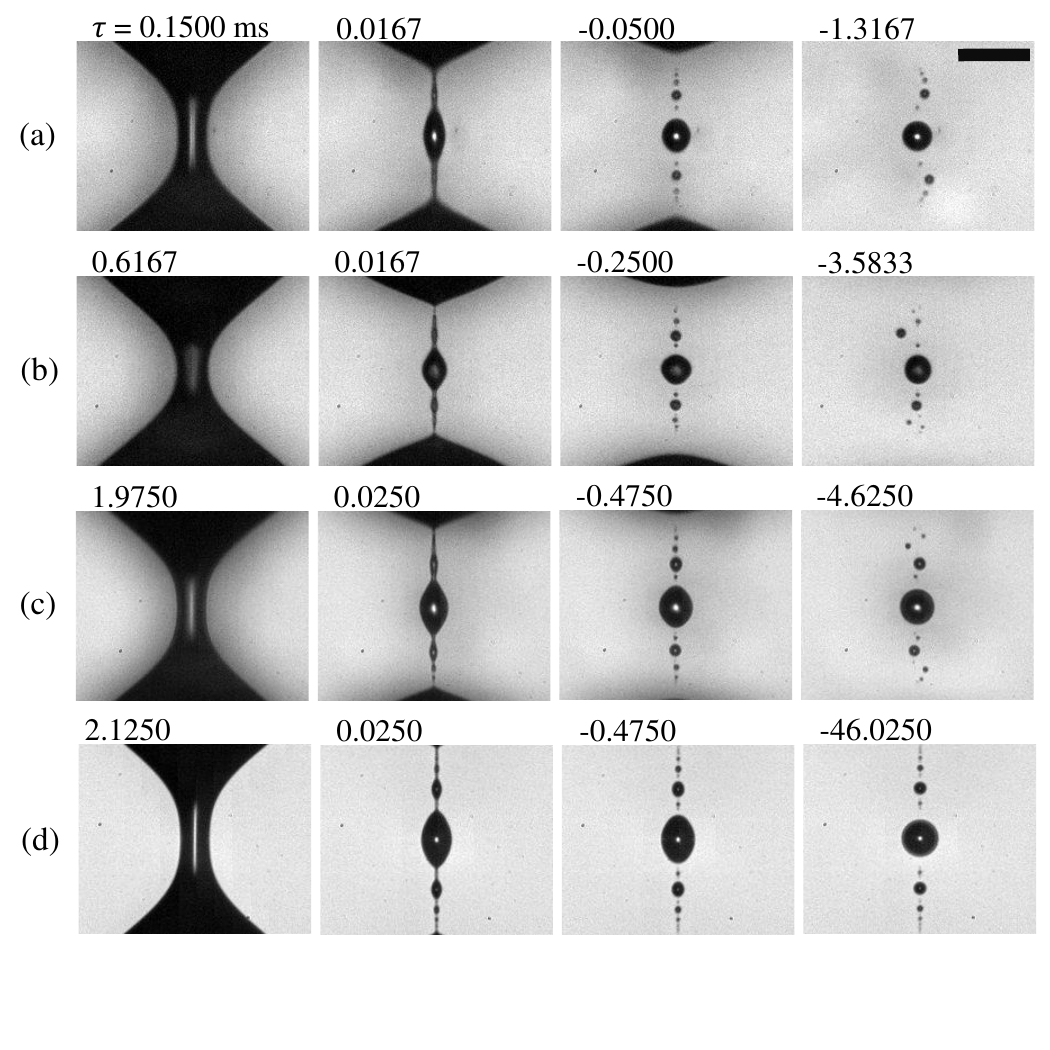}    \label{Fig2}
    \caption{Snapshots of the time evolution of liquid bridges. $\tau$ is the time before breakup and the scale bar represents \SI{100}{\micro m}. On a non-grounded, dielectric glass substrate, liquids of (a) 30\%, (b) 50\%, and (c) 70\% exhibit spontaneous and apparantly erratic motion of satellite droplets. (d) A 70\% liquid on a grounded, conductive substrate without satellite drop motion.}\label{Fig2}
\end{figure*}

During the breakup of liquid bridges on surfaces, a chain of satellite droplets forms at the symmetry plane \cite{Hartmann.2021, Farrokhi.2024}. When performing experiments with the 30\% liquid (viscosity of \SI{2.5}{mPa\cdot s}) on non-conductive glass substrates (we will call this the \emph{non-grounded case}), we observe that after the bridge breakup, these satellite droplets spontaneously move along the surface, achieving a configuration that breaks the symmetry, see \cref{Fig2}a. The motion is much slower than the bridge breakup and comes to a halt after covering distances $\approx 10-\SI{30}{\micro m}$ in $\approx 1-\SI{2}{ms}$, yielding satellite droplet velocities of the order of $0.1-\SI{1}{mm/s}$. The final distribution of satellite droplets appears largely random. However, the length scales of droplet motion and the corresponding time scales are reproducible. This behavior is not observed in the viscous wetting liquid bridge breakup on conductive silicon wafers \cite{Farrokhi.2024}.

To rule out inertial forces as a cause for this motion, we increase the viscosity of the liquid bridge by increasing the glycerol content. This reduces the characteristic velocity of the breakup and thus the inertial forces possibly acting on the satellite droplets.
\cref{Fig2}b~and~c show the breakup process for the 50\% and 70\% liquids, with viscosities of \SI{5.9}{mPa\cdot s} and \SI{23.7}{mPa\cdot s}, respectively. While the overall timescales of the breakup process increase substantially, the satellite droplets travel about the same distance before reaching their final configuration. We observe a similar scattering of droplet morphologies when we repeat an experiment with the same set of parameters. 
At these low droplet velocities below \SI{1}{mm/s}, contact angle hysteresis governs the contact line forces \cite{Li.2023kinetic}. Surface tension does not vary substantially for different glycerol contents \cite{Takamura.2012}. Consequently, if inertial forces were responsible for the droplet motion, we would expect a less pronounced motion at higher viscosities and, correspondingly, lower velocities. 

We repeat these experiments on grounded conductive substrates (\emph{grounded case}) with the same surface properties, where slide electrification is suppressed \cite{Li.2023}. \cref{Fig2}d shows an experiment under the same conditions as \cref{Fig2}c, but for the grounded case. Apparently, the grounded substrate completely suppresses the satellite droplet motion, even when recording the dynamics over a much longer time period (compare the timestamps in the rightmost panels of \cref{Fig2}c and d). We conclude that the satellite droplet motion during liquid bridge breakup on non-conductive substrates is caused by electrostatic effects.

\subsection{Slide electrification}

When drops slide along dielectric (non-conductive) hydrophobic substrates, they leave behind a (usually negative) surface charge along their path. Due to charge conservation, the drops acquire an opposite net charge \cite{Yatsuzuka.1994,Stetten.2019,Bista.2023}. This effect is called slide electrification and is caused by charge separation at the receding contact line \cite{Ratschow.2024,Bista.2024}. At the solid-liquid interface, an electric double layer is present, comprising a physically or chemically bound layer of  surface charge and a diffuse layer of countercharge of thickness $\lambda\approx1-\SI{1000}{nm}$. The thickness of the diffuse layer is called Debye length. It is an inherent property of an electrolyte solution and scales inversely with the square root of the ion concentration $c_0$ \cite{Israelachvili.2012} 
\begin{equation}
    \lambda=\sqrt{ \frac{\varepsilon_\mathrm{r} \varepsilon_0 RT}{2 z^2F^2 c_0}},\label{eq_lambda}
\end{equation} 
where $\varepsilon_\mathrm{r}$ is the liquid's dielectric constant, $\varepsilon_0$ the vacuum permittivity, $R$ the universal gas constant, $T$ the temperature, $z$ the ion valence, and $F$ the Faraday constant. 
When the surface is dewetted at the receding contact line, a part of the bound surface charge from the electric double layer stays on the dewetted surface and leaves the liquid while its countercharge remains in the liquid. The amount of deposited charge decreases with decreasing receding contact angle and with increasing dewetting velocity and accumulated counterions in the liquid, leading to a net liquid potential \cite{Ratschow.2024,Bista.2023}.
This charge separation occurs even when the liquid is grounded \cite{Ratschow.2024} 

The dewetted surface charge interacts electrostatically with the liquid. 
\cite{Li.2023} recently showed that these electrostatic interactions constitute a hitherto overlooked contribution to contact angle hysteresis, as they decrease the receding contact angle. While microscopically, the effect is caused by Maxwell stresses near the contact line, macroscopically it can be attributed to an increase in solid surface energy $\gamma_\mathrm{S}$ due to the presence of charges and their mutual electrostatic repulsion. The increase in solid surface energy scales like \cite{Li.2023}
\begin{equation}
    \Delta\gamma_\mathrm{S}=\gamma_\mathrm{S}^\mathrm{eff}-\gamma_\mathrm{S}=\frac{\sigma^2w}{2\varepsilon_0(1+\varepsilon_\mathrm{s})},\label{eq_gamma}
\end{equation} 
with the surface charge density $\sigma$, the width of the charged area corresponding to the drop width $w$, and the substrate's dielectric constant $\varepsilon_\mathrm{s}$. Drops tend to move towards areas of higher wettability, as is commonly used in microfluidic drop transport via electrowetting \cite{Cho.2003}. 

This helps explaining the satellite droplet motion in our experiments. As the liquid bridge breaks up, its receding contact line deposits surface charge, leading to a charged area of the same length scale as the initial liquid bridge. Because the dynamic contact angle and the dewetting velocity both vary during the breakup process, the deposited surface charge is not uniform. After the breakup, the satellite droplets spontaneously move due to electrostatic interactions with the charged surface and other drops,
yielding the seemingly random morphologies seen in \cref{Fig2}a-c. 

\begin{figure}
    \centering
    \includegraphics[width=7cm,page=1]{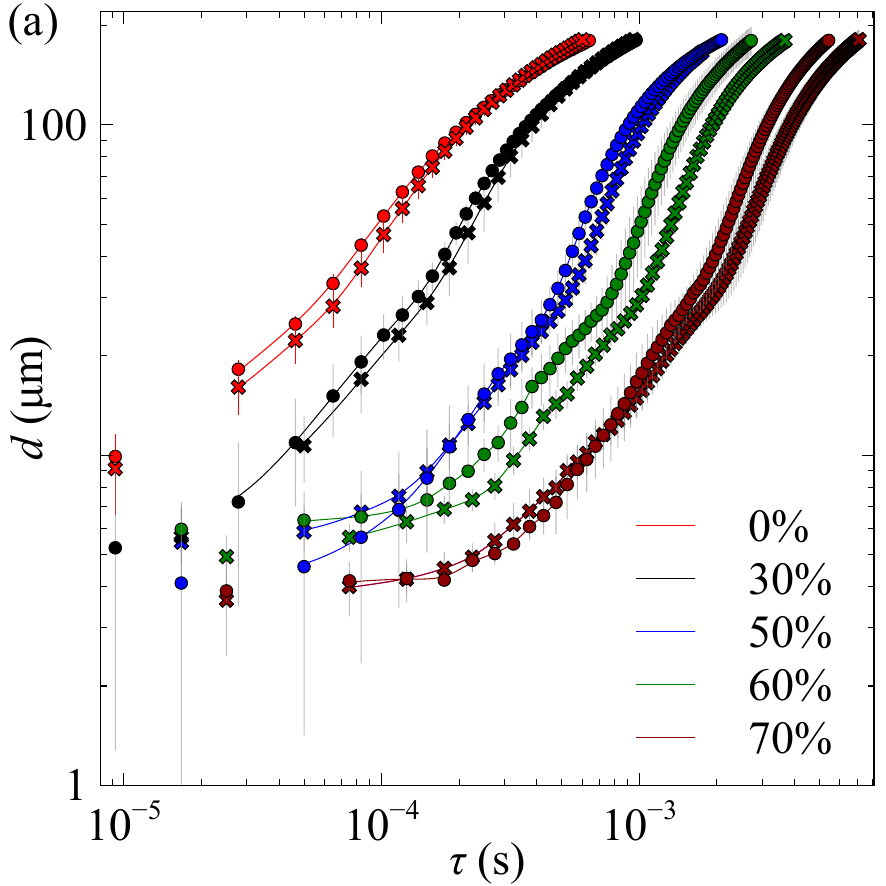}\label{Fig3a}
    \includegraphics[width=7cm,page=2]{Fig3.pdf}\label{Fig3b}
    \caption{Electrostatic influence on the breakup dynamics of liquid bridges with glycerol contents between 0\% and 70\% . Circles mark grounded conductive, and crosses non-grounded dielectric substrates. The symbols represent measurements and solid lines are B-spline fits to the data to indicate the trends. (a) Minimum width of the liquid bridge $d$ vs. time before breakup $\tau$. (b) contact line velocity $U$ at the minimum width vs. $d$ from the same data as in (a).}\label{Fig3}
\end{figure}

\subsection{Electrostatic influence on breakup dynamics}

Under the assumption that the breaking liquid bridge deposits a surface charge density $\sigma$ on a non-grounded substrate, the resulting increase in surface energy, \cref{eq_gamma}, should also influence the breakup process itself. In this section, we systematically study the influence of electrostatics on the breakup dynamics of viscous wetting liquid bridges. To this end, we keep all experimental parameters the same and only vary the substrate between grounded and non-grounded. Note that the other surface properties are dictated by the coating and do not vary between grounded and non-grounded substrates. Thus, by analyzing the difference between grounded and non-grounded substrates with otherwise identical parameters, we can identify the influence of electrostatics. A similar technique has been applied by 
\cite{Li.2022,Li.2023} with sliding drops. 

\cref{Fig3}a shows the minimum lateral width $d$ of the wetting liquid bridge against the time until breakup $\tau$. \cref{Fig3}b shows the minimum width $d$ versus the contact line velocity $U$ at the position of minimum width. Circles represent the grounded and crosses the non-grounded case, respectively. The velocities throughout the breakup process vary across one and a half orders of magnitude between the liquids of different glycerol content. When looking at the individual pairs of grounded and non-grounded cases in \cref{Fig3}b at the time of maximum velocity, there is little difference for pure water (red). For 30\% glycerol, the non-grounded case shows roughly $5-10$\% lower velocities. This difference increases with the glycerol content, amounting to about 15\% with the 50\% liquid and exceeding 30\% for the highest glycerol contents, 60\% and 70\%. Overall, the influence of electrostatics increases with the glycerol content, which in turn increases viscosity and decreases velocity. 

Because of the complexity of the bridge breakup process, we cannot quantitatively extract the additional contact line force due to electrostatics. However, we have identified a parametric dependency that we can compare to slide electrification theory in order to corroborate our hypothesis that the observed effects are caused by slide electrification. 

\subsection{Parametric trends}

The surface charge deposited due to slide electrification decreases at higher dewetting velocities \cite{Ratschow.2024}. There exists an upward flow near the receding contact line. When this flow is strong enough, it drives counterions in the diffuse layer upward (along the gas-liquid interface) and thereby increases the effective Debye length $\lambda_\mathrm{eff}$. The surface charge density that exists in chemical equilibrium with the diffuse charge can approximately be expressed by the linearized Grahame equation \cite{Ratschow.2024,Bista.2023}
\begin{equation}
\sigma=\frac{\varepsilon_\mathrm{r} \varepsilon_0 \zeta}{\lambda_\mathrm{eff}}, \label{eq_grahame} 
\end{equation}
where $\zeta$ is the zeta potential of the solid-liquid interface. Apparently, the surface charge density decreases with increasing effective Debye length. 
The relative strength of the upward flow can be measured with a Péclet number, $\mathrm{Pe}=U \lambda / D$,
where $D$ is the ion diffusivity. Finally, the effective Debye length is \cite{Ratschow.2024}
\begin{equation}
    \lambda_\mathrm{eff}=\lambda \frac{\sqrt{\mathrm{Pe}^2+4}+\mathrm{Pe} }{2}. \label{eq_lambda_eff}
\end{equation}
At low Péclet numbers $\ll1$, the influence of convection is negligible and $\lambda_\mathrm{eff} \approx \lambda$. At high Péclet numbers $\gg1$, convective transport is dominant over diffusion and the Debye length extends to $\lambda_\mathrm{eff} \approx \lambda \mathrm{Pe}$. 

Experimentally, we find that the higher the glycerol content, the larger the deviation between the grounded and non-grounded substrates and thus the stronger the electrostatic effects. 
In order to compare our experimentally observed trend with the theoretical prediction, we need to estimate the anticipated changes in surface charge density between the different glycerol content liquids according to \cref{eq_grahame,eq_lambda_eff}. 

Since the surface does not change, the zeta potential $\zeta$ should stay constant between the experiments. However, changing the glycerol content alters the Péclet number $\mathrm{Pe}$ and the effective Debye length $\lambda_\mathrm{eff}$. 
Characteristic scales for the velocity are available from the measurements shown in \cref{Fig3}b and vary by more than a factor of 10. The increasing glycerol concentration increases the liquid viscosity and slows down the process. Yet, it also affects the other relevant parameters influencing the Péclet number and the effective Debye length: the dielectric constant $\varepsilon_\mathrm{r}$ and the ion concentration $c_0$, as well as the ion diffusivity $D$. 
In the following, we analyze how these other variables vary and if the overall difference between the mixtures is dominated by velocity changes.

\subsubsection{Dielectric constant}

Water and glycerol are both polar solvents with dielectric constants at room temperature of  $\varepsilon_\mathrm{water}=78$ \cite[p. 1004]{Atkins.2006} and $\varepsilon_\mathrm{glycerol}=41$ \cite{Akerlof.1932}. The Debye length scales $\propto\sqrt{\varepsilon_\mathrm{r}}$, so the influence of the dielectric constant of pure glycerol compared to pure water yields a decrease in Debye length of just 27\%. In our experiments, we are only going up to a glycerol content of 70\% in water and thus expect an even smaller variation. 
Even for a constant effective Debye length, the higher dielectric constant of pure water would yield a slightly increased surface charge according to \cref{eq_grahame}, opposite to the experimentally observed trend.
Yet, these influences on the 
surface charge density
by themselves are small compared to the influence due to velocity changes through the Péclet number. 

\subsubsection{Ion concentration}

While we do not add any ions to the liquid, the autodissociation of water into protons and hydroxide ions leads to the presence of ions even in ultrapure water. The $p\mathrm{K_A}$ of this reaction is 14 \cite{Silverstein.2017}. Glycerol is a propane molecule with one hydroxyl group attached to each carbon atom. Similar to water, the hydroxyl groups in glycerol can dissociate and release protons. This reaction has a $p\mathrm{K_A}$ of 14.4 \cite{Perrin.1984}, which is very similar to water. Thus, the ion concentrations of ultrapure water and glycerol due to dissociation are essentially the same and should not vary significantly between the mixtures. The theoretical Debye length of water due to autodissociation is $\lambda=\SI{971}{nm}$.

However, we conduct our experiments at ambient conditions in air. \cite{vogel} showed that in this case, the dissolution of atmospheric 
\ce{CO2} in water is non-negligible and increases the ion concentration, decreasing the Debye length. At ambient conditions, the solubility of \ce{CO2} in water is 
$\approx 7 \, \mathrm{mol} / \mathrm{kmol}$ \cite{Dodds.1956}. In comparison, the solubility of \ce{CO2} in pure glycerol at ambient conditions is $\approx 2 \, \mathrm{mol} / \mathrm{kmol}$ \cite{Nunes.2013}. Accordingly, the ion concentration in water under atmospheric conditions can be expected to be slightly higher than in pure glycerol. This would yield a Debye length that is slightly higher in pure glycerol by a factor of $\approx1.87$.  

Overall, we conclude that the Debye length of our liquids does not vary substantially with the glycerol content 
and use the value for water in contact with atmospheric \ce{CO2} at $p\mathrm{H}=5.5$ \cite{vogel}, $\lambda \approx \SI{170}{nm}$, for our estimations of the Péclet number and the changes in surface charge density.

\subsubsection{Ion diffusivity}

PDMS coated glass surfaces acquire a negative charge during slide electrification \cite{Stetten.2019,Wong.2022,Leibauer.2024}. Thus, the only available counterions in our liquids are protons, which accumulate in the diffuse layer. The diffusion of protons in water is governed by the Grotthus mechanism, a proton hopping mechanism yielding diffusivities much larger than conventional diffusion \cite{Grotthuss.1806,Agmon.1995}. Due to the abundance of hydroxyl groups, the Grotthus mechanism also dominates in pure glycerol \cite{Zhuang.2004}. Thus, despite the higher viscosity of glycerol compared to water, the diffusivity of protons is comparable between the two liquids. 

\subsubsection{Péclet numbers and charge separation in glycerol-water mixtures}

The above analysis has important implications. It shows that using mixtures of glycerol and water enables slide electrification experiments with varying viscosity, while keeping all other parameters essentially constant. 

Here, we estimate the Péclet numbers in our experiments based on the properties of pure water. 
Using $D= \SI{5e-9}{m^2/s}$ \cite{Atkins.2006} and the respective maximum velocities in \cref{Fig3}b of \SI{0.5}{m/s}, \SI{0.3}{m/s}, \SI{0.15}{m/s}, \SI{0.09}{m/s}, and \SI{0.04}{m/s},  we find Péclet numbers of
17, 10.2, 5.1, 3.0, and 1.36
for the 0\%, 30\%, 50\%, 60\%, and 70\% liquids, respectively. 

\cref{eq_grahame,eq_lambda_eff} predict that the deposited surface charge density relative to that of a contact line moving with negligible velocity is 0.059 for the 0\% liquid and 0.53 for the 70\% liquid. 
This aligns with the experimentally observed increase of electrostatic effects with increasing glycerol content, and correspondingly, increased viscosity. 
In total, this analysis corroborates that slide electrification plays an important role in the breakup of wetting liquid bridges on non-conductive substrates.

\section{Conclusions}

We have demonstrated that electrostatics affects the breakup of wetting liquid bridges on non-conductive substrates. During the breakup, the contact line can be slowed down by over 30\% compared to grounded conductive surfaces. After the breakup, satellite droplets spontaneously move along the surface for tens of micrometers before reaching random final deposition morphologies. The influence on bridge breakup increases with the liquid viscosity. We found that this trend is in accordance with slide electrification theory. Up to now, slide electrification effects have been identified for drops sliding along surfaces. Our findings indicate that the relevance of slide electrification goes beyond such scenarios and suggest that it is a common phenomenon occurring in dynamic wetting.

\begin{acknowledgments}
We thank G\"unter K. Auernhammer and Peyman Rostami for insightful discussions, and Mohammad Ali Hormozi for the AFM characterization of our substrates.
This work was supported by the German Research Foundation (DFG) within the Collaborative Research Centre 1194 “Interaction of Transport and Wetting Processes,” Project ID No. 265191195, subproject A02b.

S.J.F. devised, designed, conducted, and evaluated the experiments, A.D.R. developed the theoretical framework, provided the interpretation of the results, and wrote the manuscript, S.H. proposed and supervised the work. S.J.F. and A.D.R. contributed equally.
\end{acknowledgments}


%

\end{document}